\documentclass{ragtime}

\usepackage{amsthm}
\usepackage[mathscr]{euscript}
\usepackage{times}
\usepackage{txfonts}
\usepackage{color}

\providecommand{\dif}{\mathrm{d}}
\providecommand{\en}{\mathscr{E}}
\providecommand{\pot}{\mathscr{U}}
\providecommand{\aaa}{\mathscr{A}}

\graphicspath{{mat/}}

\title[Oscillations of tori in the vicinity of neutron stars]
{
	Oscillations of non-slender tori in the external Hartle-Thorne geometry
}

\author[M. Matuszkov\'a et al.]
{Monika Matuszkov\'a\at{1,a}
	Kate\v{r}ina Klimovi\v{c}ov\'a\at{1,b}
	\splitauthors Gabriela Urbancov\'a\at{1,c}
	Debora Lan\v{c}ov\'a\at{1,d}  \splitauthors
	Eva \v{S}r\'{a}mkov\'{a}\at[]{1}
	and Gabriel T\"{o}r\"{o}k\at[]{1}
	\\ 
	\ins{1}Research Centre for Computational Physics and Data Processing, \splitins[1]Institute of Physics,
	Silesian University in Opava, 
	Bezru\v{c}ovo n\'am.~13,\splitins[1]CZ-746\,01 Opava,
	Czech Republic\\
	\ins{a}\Email{monika.matuszkova@physics.slu.cz}\\ 
	\ins{b}\Email{katerina.klimovicova@physics.slu.cz}\\
	\ins{c}\Email{gabriela.urbancova@physics.slu.cz}\\  
	\ins{d}\Email{debora.lancova@physics.slu.cz} } 

\graphicspath{{mat/}}

\begin{document}
	
	\begin{abstract}
		We examine the influence of the quadrupole moment of a slowly rotating neutron star on the oscillations of non-slender accretion tori. We apply previously developed methods to perform analytical calculations of frequencies of the radial epicyclic mode of a torus in the specific case of the Hartle-Thorne geometry. We present here our preliminary results and provide a brief comparison between the calculated frequencies and the frequencies previously obtained assuming both standard and linearized Kerr geometry. 
		Finally, we shortly discuss the consequences for models of high-frequency quasi-periodic oscillations observed in low-mass X-ray binaries.
	\end{abstract}
	
	
	\section{Introduction}
	\label{sec:intro}

	Numerous interesting features have been discovered during the long history of X-ray observations of low-mass X-ray binaries (LMXBs). One of them is the fact that variability of the X-ray radiation coming from these sources occurs at frequencies in the order of up to hundreds of Hertz with the highest values reaching above 1.2 kHz. Even though the discovery of this rapid variability was made almost 30 years ago, to this day, there is no convincing explanation of its origin. The  phenomenon is called the high-frequency quasi-periodic oscillations (HF QPOs) and many models have been proposed in the attempt to explain its nature \citep[see, e.g.,][ and references therein]{tor-etal:2016:MNRAS,kot-etal:2020:AA}. 
	
	It has been noticed that the HF QPOs frequencies are in the same order as those corresponding to orbital motion in the very close vicinity of a compact object, such as neutron star (NS) or black hole (BH). This suggests that there is a relation between the QPO phenomenon and the physics behind the motion of matter close to the accreting object. Since positions of specific orbits in the accretion disk (such as its inner edge) and the associated orbital frequencies depend on the properties of the central object, there is a believe that it is possible to infer the compact object properties from the QPOs data.\footnote{We often use the shorter term "QPOs" instead of "HF QPOs" throughout the paper.}

	In the above context, several studies have focused on a possible relation between the QPOs and an oscillatory motion of an accretion torus formed in the innermost accretion region \citep[][]{Klu-etal:2001, Klu-etal:2004, abr-bul-etal:2003,abr-etal:2003,rez-etal:2003,Bur:2005,tor-etal:2005,don-etal:2011,tor-etal:2016:MNRAS,ave-etal:2018}.
	
	\citet{sramkova2009} and \citet{Fragile2016} have performed calculations of frequencies of the epicyclic oscillations of fluid tori assuming Kerr geometry, which describes rotating BHs. Here we follow their approach and consider slowly rotating NSs and their spacetimes described by the Hartle-Thorne geometry \citep{Har:1967:ApJ:,Har-Tho:1968:ApJ:}. We present the first, preliminary results of our calculations of the radial epicyclic oscillation frequencies and provide a brief comparison of these to the frequencies obtained previously for the Kerr and linearized Kerr geometries. Finally, we discuss some consequences for models of NS QPOs.


	


	
	\section{Oscillations of tori in axially symmetric spacetimes}
	
	We consider an axially symmetric geometry. The spacetime element may be expressed in the general form as
	\begin{align}
		\label{eq:stel}
		\dif s^2 = g_{tt} \dif t^2 + 2 g_{t\varphi} \dif t \dif \varphi + g_{rr} \dif r^2 + g_{\theta\theta} \dif \theta^2 + g_{\varphi\varphi} \dif \varphi^2.  
	\end{align}
	We use the units in which $c = G = 1$ with $c$ being the speed of light and $G$ the gravitational constant.
	
	\subsection{Equilibrium configuration} 
	
	We assume a perfect fluid torus in the state of pure rotation with constant specific angular momentum~$l$ as described in  \cite{abramowicz2006,blaes2006}. 
	
	In this case, the fluid forming the torus has a four-velocity $u^\mu$ with only two non-zero components,
	\begin{equation}
		u^\mu = A(1,0,0,\Omega),
	\end{equation}
	where $A$ is the time component $u^{t}$ and $\Omega$ is the orbital velocity. One may write 
	\begin{align}
		A =& u^{t} =  ( - g_{tt} - 2 \Omega g_{t\varphi} - \Omega^2 g_{\varphi\varphi} )^{-1/2}, \\
		\Omega =& \dfrac{u^\varphi}{u^t} = \frac{g^{t\varphi} - l g^{\varphi\varphi}}{g^{tt} - l g^{t\varphi}}.
	\end{align}
	The perfect fluid with density $\rho$, pressure $p$ and the energy density $e$ is characterised by the stress-energy tensor
	\begin{equation}
		T^{\mu\nu} = (p+e) u^\mu u^\nu + p g^{\mu\nu}.
	\end{equation}
	For a polytropic fluid, we may write: 
	\begin{align}
		\label{pol}
		p &= K \rho^\frac{n+1}{n},\\
		e &= np + \rho,
	\end{align}
	where $K$ and $n$ denote the polytropic constant and the polytropic index, respectively. In this work, we use $n=3$, which describes a radiation-pressure-dominated torus.
	
	The Euler formula is obtained from the energy–momentum  conservation law, $ \nabla_\mu T^\mu_{\; \ \; \nu} = 0 $, using the assumption of $ l=const. $~\citep{abramowicz1978,abramowicz2006} 
	\begin{equation}
		\label{eul}
		\nabla_\mu (\ln \en) = - \frac{\nabla_\mu p}{p + e}, 
	\end{equation}
	with  $\en$  being the specific energy 
	\begin{equation}
		\en = - u_t = \left( - g^{tt} + 2 l g^{t\varphi} - l^2 g^{\varphi\varphi} \right)^{-1/2}.
	\end{equation}
	
	By integrating (\ref{eul}) we obtain the Bernoulli equation ~\citep{Fragile2016,Horak2017} 
	\begin{equation}
		\label{ber}
		H \en = const.,
	\end{equation}
	where $H = \frac{p+e}{\rho}$ denotes the enthalpy in the form presented by \cite{Fragile2016} and \cite{Horak2017}. 
	From relation (\ref{ber}), we can derive the equations describing the structure and shape of the torus: 
	\begin{align}
		\frac{p}{\rho}  &= \frac{p_0}{\rho_0} f( r, \theta ), \label{pkuro} \\
		f( r, \theta ) &= \frac{1}{nc_{\mathrm{s},0}^2} \left[ \left( 1 + nc_{\mathrm{s},0}^2 \right) \frac{\en_0}{\en} - 1 \right], \label{f} \\
	\end{align}
	where $ c_{\mathrm{s}} $ is the sound speed in the fluid defined as~\citep{abramowicz2006}\footnote{The definition is fully valid for $c_{\mathrm{s}} << 1$, but this has no significant effect on our results.} 
	\begin{equation}
		c_{\mathrm{s}}^2 =  \frac{\partial p}{\partial \rho}  = \frac{n+1}{n} \frac{p}{\rho},
	\end{equation}
	and the subscript $0$ denotes the quantities evaluated at the torus centre.  
	From equations (\ref{pol}) and (\ref{pkuro}), one can obtain the following formulae for pressure and density of the fluid:
	\begin{align}
		p &= p_0 \left [ f \left( r,\theta \right ) \right ]^{n+1},  \\ 
		\rho &= \rho _0 \left [ f \left( r,\theta \right ) \right ]^n.
	\end{align}
	
	It is useful to introduce new coordinates $\overline{x}$ and $\overline{y}$ by relations
	
	\begin{align}
		\overline{x} =& \frac{\sqrt{ g_{rr,0} }}{\beta} \left( \frac{r - r_0}{r_0} \right), \\ 
		\overline{y} =& \frac{\sqrt{g_{\theta\theta,0}}}{\beta} \left( \frac{\frac{\pi}{2} - \theta}{r_0} \right).
	\end{align}
	In these coordinates, we have $\overline{x}=0$ and $\overline{y} = 0$ at the torus centre. We furthermore introduce a $\beta$ parameter determining the torus thickness, which is connected to the sound speed at the torus centre in the following manner \citep{abramowicz2006,blaes2006}:
	\begin{align}
		\beta^2 =& \frac{2nc_{\mathrm{S},0}^2}{r_0^2 \Omega_0^2 A_0^2}.
	\end{align}
	
\begin{figure}[ht]
		\centering
		
		\includegraphics[width=0.5\linewidth]{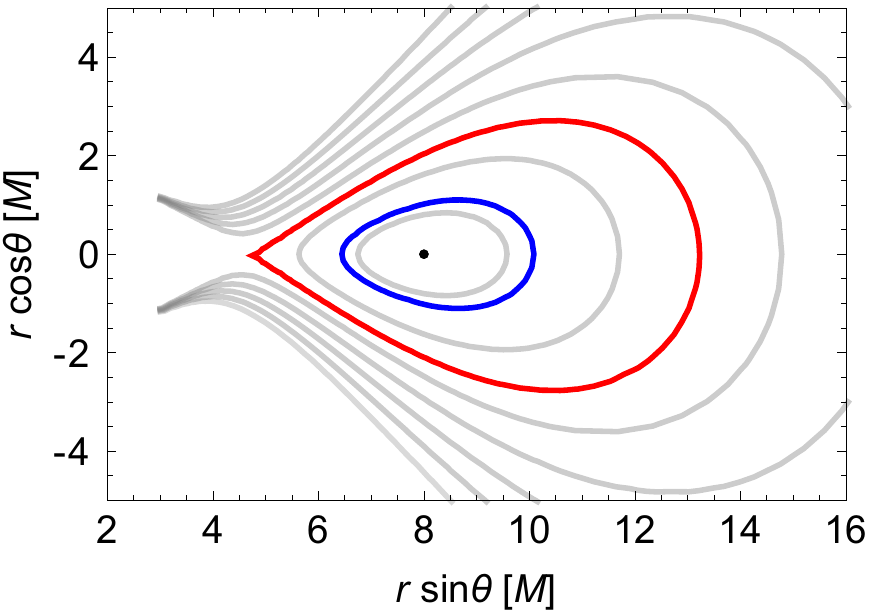}
		\caption{Meridional cross-section illustrating the shape of the equipressure surfaces in the Schwarzschild geometry. The red line marks a cusp torus with $\beta = \beta_{\mathrm{cusp}}$, the blue line corresponds to an equilibrium torus with $\beta < \beta_{\mathrm{cusp}}$, and the black dot denotes the centre of the torus (as well as the infinitely slender torus with $\beta \rightarrow 0$). 
		}
		\label{torustem}
	\end{figure}

	The surface of the torus, which coincides with the surface of zero pressure, is given by the condition $f(r, \theta ) =0$. 
	An example of the torus cross-section is shown in Figure \ref{torustem} illustrating the character of the equipressure surfaces for different values of $\beta$. An equilibrium torus is formed when the perfect fluid fills up a closed equipressure surface. The largest possible torus arises by filling up the equipressure surface that has a crossing point -- the so-called cusp. We call this structure, for which we have $\beta = \beta_{\mathrm{cusp}}$, the "cusp torus". Notice that, for $\beta > \beta_{\mathrm{cusp}}$, the equipressure surfaces are no longer closed and no torus therefore can be formed.

	\subsection{The oscillating configuration}
	
	We assume the effective potential $\pot$ \citep[e.g.][]{abramowicz2006} in the form 
	\begin{equation}
		\pot = g^{tt} - 2 l_0 g^{t\varphi} +  l_0^2 g^{\varphi\varphi}.
	\end{equation}
	An infinitesimally slender torus with $\beta\rightarrow0$ at $r_0$ with specific angular momentum  $l_0$ undergoing a small axially symmetric perturbation in the radial direction 
	will oscillate with the frequency equal to the radial epicyclic frequency of a free test particle given by ~\citep{abramowicz2005,aliev1981}
	\begin{equation}
		\label{epeq}
		\nu_{r}^2 = \left. \frac{1}{4\pi^2} \frac{\en_{0}^2}{ 2 A_{0}^2 g_{rr,0}} \frac{\partial^2 \pot}{\partial r^2}\right\vert_{0}.
	\end{equation}
	
	Now let us investigate how the frequency changes when the torus becomes thicker and/or when the perturbation is not axially symmetric. 
	Assume small perturbations of all quantities around the equilibrium state in the form~\citep{abramowicz2006,blaes2006}
	\begin{equation}
		\delta X (t, r, \theta, \varphi) = \delta X (r, \theta) \mathrm{e}^{i \left( m \varphi - \omega t \right)},
	\end{equation}
	where $ m $ is the azimuthal number and $ \omega $ is the angular frequency of the oscillations. In this work, we focus on two modes of oscillations: the axially symmetric ($m=0$) and the first non-axisymmetric ($m=-1$) radial epicyclic modes.

	From the continuity equation
	$ \nabla_\mu \left( \rho u^\mu \right) = 0$,
	one can get the relativistic version of the Papaloizou-Pringle equation~\citep{abramowicz2006,Fragile2016}, \footnote{For the sake of simplicity, from now on, we will use $f = f(r,\theta)$.}
	\begin{align} 
		\frac{1}{\sqrt{-g}} \partial _\mu \frac{ \sqrt{-g} g^{\mu\nu} f^n \partial_\nu W }{nc_{\mathrm{s},0}^2 f +1}
		+& \left( l_0 \omega - m \right)^2 
		\frac{\Omega g^{t\phi} -g^{\phi \phi}}{1-\Omega l_0}
		\frac{f^{n}}{nc_{\mathrm{s},0}^2 f + 1}
		W 	= \nonumber \\ 
		=&  - \frac{2 n \aaa^2 \left( \overline{\omega} - m \overline{\Omega} \right) ^2}{\beta^2 r_0^2} f^{n-1} W, 
		\label{papa}
	\end{align}
	where $ \lbrace \mu,\nu \rbrace \in \lbrace r, \theta \rbrace $, $ \aaa \equiv A/A_0 $, 
	$ \overline{\Omega} \equiv \Omega / \Omega_0 $,   
	$ \overline{\omega} \equiv \omega / \Omega_0 $, $ g $ is the determinant of the metric tensor and $W$ equals to~\citep{abramowicz2006}
	\begin{equation}
		W = - \frac{\delta p}{A \rho \left( \omega - m \Omega \right)}.
	\end{equation}
	
	Equation (\ref{papa}) has no analytical solution except for the limit case of an infinitely slender torus $ \left( \beta \rightarrow 0 \right) $.
	In the case of non-slender tori  $ \left( \beta > 0 \right)$, the equation can be solved using a perturbation method (see, e.g., \cite{sramkova2009}).
	
	\subsubsection{Solving the Papaloizou-Pringle equation}
	When the exact solution for a simplified case is known (as for $ \beta \rightarrow 0 $), we can use perturbation theory to find the solution for more complicated cases ($ \beta > 0 $). \footnote{Note the perturbation method gives reasonable results only for small values of $\beta$ and our results are therefore valid only for slightly non-slender tori.}
	
	
	By expanding the quantities $ \overline{\omega} $, $ W $, $\aaa$, $\overline{\Omega}$, $f $ in $ \beta $ \citep{sramkova2009}
	\begin{equation}
		Q = Q^{(0)} + \beta Q^{(1)} + \beta^2 Q^{(2)} + \cdots, \qquad
		Q \in \left\lbrace \overline{\omega}, W, \aaa, \overline{\Omega}, f \right\rbrace,
		\label{rozvoj}
	\end{equation}
	substituting that into equation (\ref{papa}), and comparing the coefficients of appropriate order in $ \beta $, we obtain the corresponding corrections to $W$ and $\omega$. Note that the zero order corresponds to the slender torus case ($\beta \rightarrow 0$), in which we have $\omega =  2 \pi \nu_{r}$.

	Using this procedure, \citet{sramkova2009} have derived the expression for the radial epicyclic mode frequency with the second order accuracy, which may be written as
	\begin{equation}
		\label{radeq}
		\omega_{r,\mathrm{m}} = 2 \pi \, \nu_{r} + m \, \Omega_0
		+ P_{\mathrm{m}}\,  \beta^2 + \mathscr{O} \left( \beta^3 \right),
	\end{equation}
	where $P_{\mathrm{m}}$ denotes the second order correction term for which the explicit form can be found in their paper.
	
	\section{The Hartle-Thorne geometry}\label{ht}
	
	The exterior solution of the Hartle-Thorne metric is characterized by three parameters: the gravitational mass $M$, angular momentum $J$ and the quadrupole moment $Q$ of the star. We use this metric assuming dimensionless forms of the angular momentum and the quadrupole moment, $j=J/M^2$ and $q=Q/M^3$, which can be in the Schwarzschild coordinates written as \citep{Abramowicz2003}\footnote{Note misprints in the original paper.}:
	\begin{align}
		g_{tt} &= -\left( 1- \frac{2M}{r} \right) \left[ 1 + j^2 F_1 (r) + q F_2 (r) \right], \\ 
		g_{rr} &=  \left( 1- \frac{2M}{r} \right)^{-1} \left[ 1 + j^2 G_1 (r) - q F_2 (r) \right] , \\ 
		g_{\theta\theta} &=  r^2 \left[ 1 + j^2 H_1 (r) + q H_2 (r) \right] , \\ 
		g_{\varphi\varphi} &=  r^2 \sin^2 \theta \left[ 1 + j^2 H_1 (r) + q H_2 (r) \right] , \\
		g_{t\varphi} &=  - \frac{2M^2}{r} j \sin^2 \theta , 
	\end{align}
	where (using the $u = \cos \theta$ substitution)
	\begin{align} \nonumber
		F_1(r) = & - \left[ 8Mr^4(r-2M) \right]^{-1} \\ \nonumber
		&\left[ u^2 \left( 48M^6 - 8M^5r - 24M^4r^2 - 30M^3r^3 - 60M^2r^4 + 135Mr^5 - 45r^6 \right) \right. \\ 
		& +  \left.(r-M) \left( 16M^5 + 8M^4r - 10M^2r^3 - 30Mr^4 + 15r^5 \right) \right] + A_1(r), 
	\end{align}
	\begin{align}
		F_2(r) = & \left[ 8Mr (r-2M) \right]^{-1} 
		\left[ 5 \left(3u^2 - 1\right) \left(r-M\right) \left( 2M^2 + 6Mr - 3r^2 \right) \right] - A_1(r), \\ 
		G_1(r) = & \left[ 8Mr (r-2M) \right]^{-1} 
		\left[ \left( L(r) - 72M^5r \right) - 3u^2 \left( L(r) - 56M^5r \right) \right] - A_1(r), \\ 
		L(r) = & 80 M^6 + 8M^4r^2 + 10M^3r^3 + 20M^2r^4 - 45Mr^5 + 15r^6, \\
		A_1(r) = & \frac{15 \left(r^2-2M\right) \left(1-3u^2\right)}{16M^2} \ln \left( \frac{r}{r-2M} \right), \\ \nonumber
		H_1(r) = & \left( 8Mr^4 \right)^{-1} \left( 1-3u^2 \right) \left( 16M^5 + 8M^4r - 10M^2r^3 + 15Mr^4 + 15r^5 \right)  \\  & + A_2(r), \\
		H_2(r) = & \left(8Mr\right)^{-1} 5 \left(1-3u^2\right) \left( 2M^2 - 3Mr - 3r^2 \right) - A_2(r), \\ 
		A_2(r) = & \frac{15 \left(r^2-2M\right) \left(3u^2-1\right)}{16M^2} \ln \left( \frac{r}{r-2M} \right).
	\end{align}
	
	While for $j=0$ and $q=0$ the Hartle-Thorne metric coincides with the Schwarzschild metric, by setting $j=a/M$ and $q=j^2$ and performing a coordinate transformation into the Boyer-Lindquist coordinates \citep{Abramowicz2003},
	\begin{align}
		r_{\mathrm{BL}} &= r - \frac{a^2}{2r^3} \left[ (r+2M) (r-2M) + u^2 (r-2M) (r+3M) \right], \\ 
		\theta_{\mathrm{BL}} &= \theta - \frac{a^2}{2r^3} (r+2M) \cos \theta \sin \theta,
	\end{align}
	we obtain Kerr geometry expanded upon the second order in the dimensionless angular momentum.

	
	\section{Oscillations of tori in the vicinity of rotating neutron stars}
	
	Let us now study the changes that arise in the torus structure and for the frequencies of its oscillations when the Hartle-Thorne geometry is assumed to describe the spacetime geometry.\footnote{Following \cite{sramkova2009} and \cite{Fragile2016}, we use a Wolfram Mathematica code, which has been extended to the Hartle-Thorne geometry.} The main motivation behind this analysis is related to models of NS QPOs. While the Kerr geometry is (likely) proper to be used in the context of BH QPOs \citep[e.g., ][]{kot-etal:2020:AA}, its validity in the case of NS QPOs is limited to very compact NSs only. 
	
	\subsection{The Hartle-Thorne geometry parameters range relevant to rotating NSs}
	
	A thorough discussion of the relevance of the Hartle-Thorne geometry for the calculations of the geodesic orbital motion and QPO models frequencies is presented in \cite{Urbancova2019}. Here we just briefly summarize the appropriate ranges of the individual parameters that are implied by the present NS equations of state. The maximum value of the specific angular momentum of a NS is about  $j_{\mathrm{max}}~\sim~0.7$, the specific quadrupole moment takes values from $q/j^2~\sim~1.5$ for a very massive (compact) NS up to $q/j^2~\sim~10$ for a low-mass NS \citep{Urbancova2019}. The conservative expectations of the NS mass values are about $1.4-2.5\,M_\odot$.

	\begin{figure}[ht]
		\begin{center}
			\includegraphics[width=0.8\linewidth]{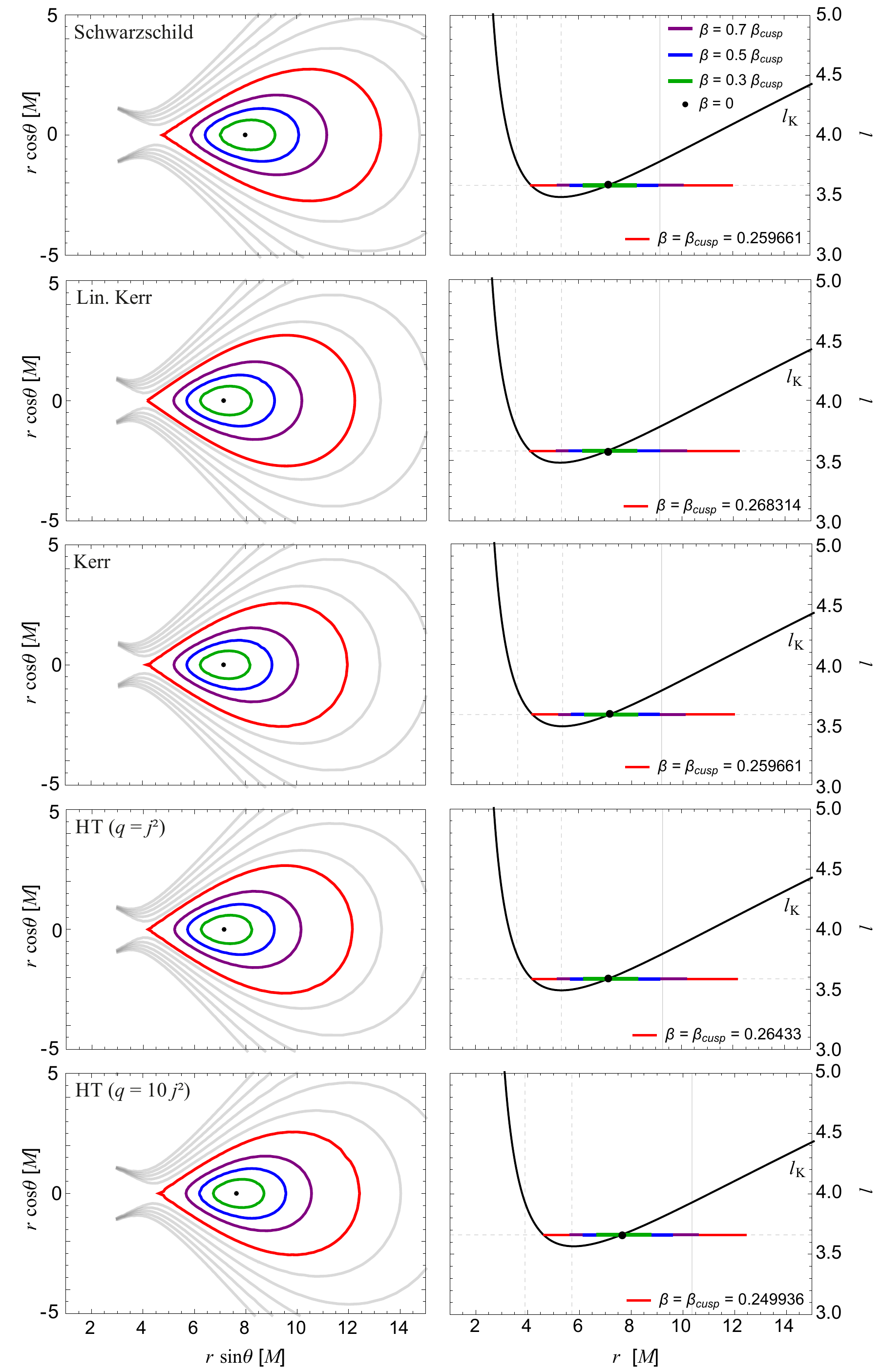}
			\caption{Illustration of some characteristics of tori carried out in different geometries. The tori are centered at the radial coordinate at which the radial epicyclic frequency of a free test particle reaches its maximum. Left panels: Meridional cross-sections of the equipressure surfaces determining the shape of the tori. From top to bottom, the results correspond to calculations carried out in the Schwarzschild, linearized Kerr ($j = 0.2$), Kerr ($j = 0.2$), and the Hartle-Thorne ($j = 0.2, q = j^2$ and $j = 0.2, q = 10 j^2$) geometry.  Right panels: Plots of the specific angular momentum of the fluid (which is constant across the torus) along with the Keplerian angular momentum. The intersection points of the two functions marked by the spots correspond to the centre of the tori. The coloured segments indicate the corresponding radial extentions of the tori. 
				\label{povrch1}}
		\end{center}
	\end{figure}
	
	\begin{figure}[ht]
		\begin{center}
			\includegraphics[width=0.8\linewidth]{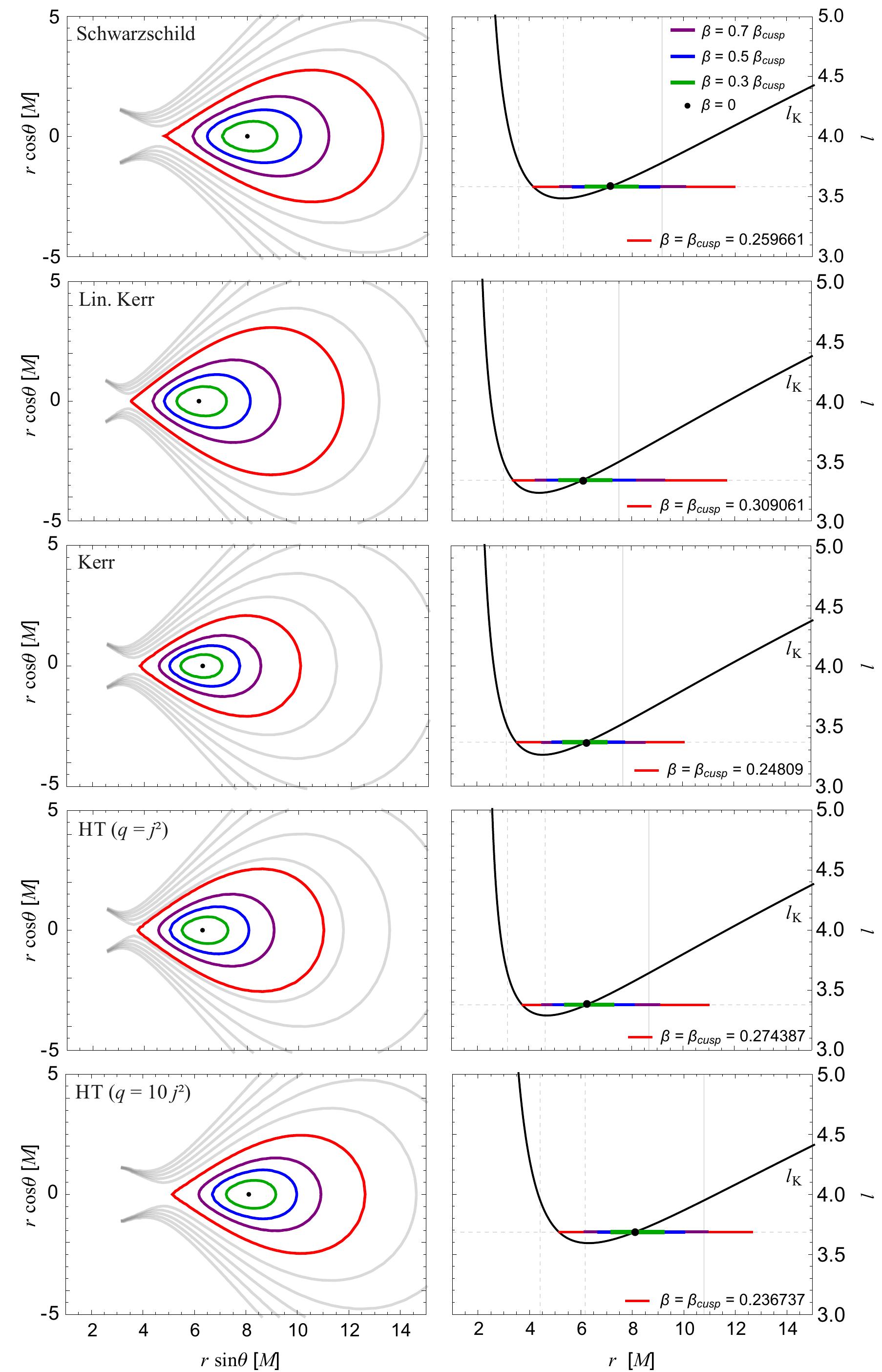}
			\caption{The same as in Figure \ref{povrch1} but for $j = 0.4$. \label{povrch2} }
		\end{center}
	\end{figure}

	\subsection{The quadrupole moment influence on the non-oscillating torus shape and size}
	
	In Figures~\ref{povrch1}~and~\ref{povrch2}, we present meridional cross-sections of tori carried out in different geometries, namely the Schwarzschild, Kerr, linearized Kerr, and the Hatle-Thorne geometry. The figures also show plots of the Keplerian angular momentum and the angular momentum of the fluid (which is constant across the torus), and the radial extentions of the tori. 
	For both figures, the top panels correspond to $j=0$ (a non-rotating NS, i.e., the Schwarzschild geometry), and the bottom panels to $j=0.2$ (Figure~\ref{povrch1}) and $j=0.4$ (Figure~\ref{povrch2}). The radial coordinate $r_0$ is chosen such that the radial epicyclic frequency of a free test particle defined at this coordinate reaches its maximum.
	
	In Table~\ref{tab:geo}, we provide a quantitative comparison of the radial extensions of tori from Figures ~\ref{povrch1}~and~\ref{povrch2}. It is given in terms of the proper radial distance, $r_{\mathrm{prop}}$, measured between the minimal, $r_{\mathrm{min}}$, and the maximal, $r_{\mathrm{max}}$, radial coordinate of the torus surface,
	\begin{equation}
		\Delta r_{\mathrm{prop}} = \int^{r_{\mathrm{max}}}_{r_{\mathrm{min}}} \sqrt{g_{rr}} \,\, \mathrm{d} r.
	\end{equation}
	
	\begin{table}[ht]
		\caption{The percentual differences in the proper radial extension $\Delta r_{\mathrm{prop}}$ of tori in the Hartle-Thorne geometry and in the Schwarzschild, Kerr, and linearized Kerr geometries.  The displayed values correspond to the situations illustrated in Figures \ref{povrch1} and \ref{povrch2}.} 
		\begin{center}
			\renewcommand{\arraystretch}{1.3}
			\begin{tabular}{lcccccc}\hline \hline
				Geometry & \multicolumn{2}{c}{Schwarzschild}   & \multicolumn{2}{c}{Kerr} & \multicolumn{2}{c}{Lin. Kerr}  \\
				\hline
				Spin $j$ & $0.2$ & $0.4$ & $0.2$ & $0.4$ & $0.2$ & $0.4$ \\
				\hline
				HT ($q = j^2$)  & $ -\,5\,\%$ & $ -\,11\,\%$	& $  +\,2\,\%$ & $ -\,5\,\%$ &  	$- \,1\,\%$ & 	$- \,13\,\%$ \\
				HT ($ q = 10 j^2$) &  $- \,7\,\%$ & $ - \,12\,\%$& 	 $- \,1\,\%$ & $-\,6\,\%$	& $- \,4\,\%$ & $- \,14\,\%$  \\
				\hline
				\hline
			\end{tabular}
		\end{center}
		\label{tab:geo}
	\end{table}

	\subsection{The quadrupole moment influence on the radial epicyclic oscillations of non-slender tori}
	
\begin{figure}[ht]
		\begin{center}
			\includegraphics[width=0.85\linewidth]{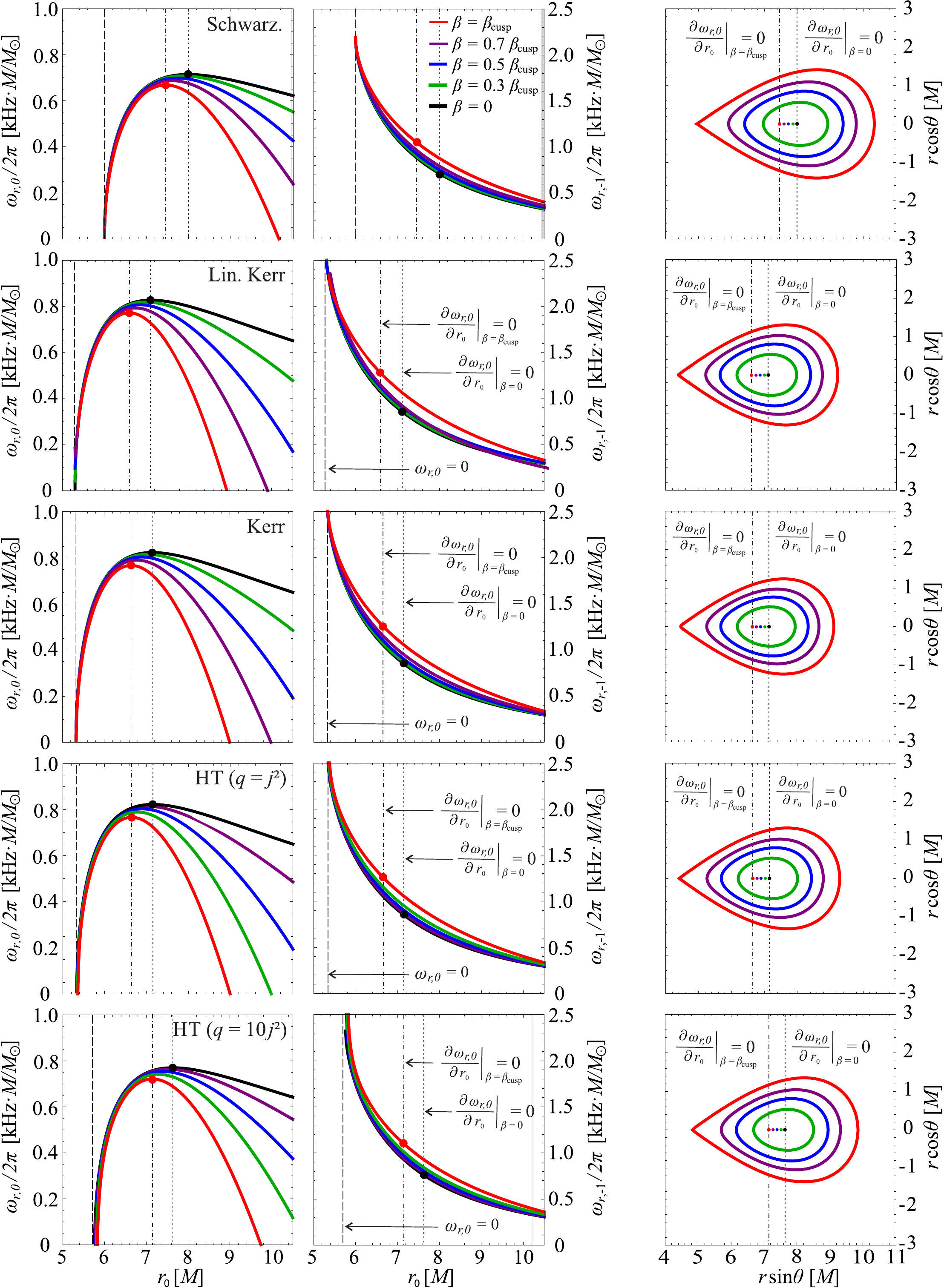}
			\caption{Frequencies of the radial epicyclic mode.  
				\newline Left panels: The $m=0$ case. From top to bottom: the Schwarzschild, linearized Kerr, Kerr, and the Hartle-Thorne ($q = j^2$ and $q = 10 j^2$) geometry. For rotating stars, we assume $j = 0.2$. The maximal frequencies allowed for the slender torus and for the cusp torus are denoted by the black and red spots, respectively.
				\newline Middle panels: The same but for the $m = -1$ case. The coloured spots denote the frequency value corresponding to the radius at which the $m=0$ radial mode frequency has its maximum.
				\newline Right panels: Tori that would oscillate with the maximal value of the $m=0$ radial epicyclic mode frequency for a given torus thickness.}	
			\label{j_0_2} 
		\end{center}
	\end{figure}

\begin{figure}[ht]
		
		\begin{center}
			\includegraphics[width=0.85\linewidth]{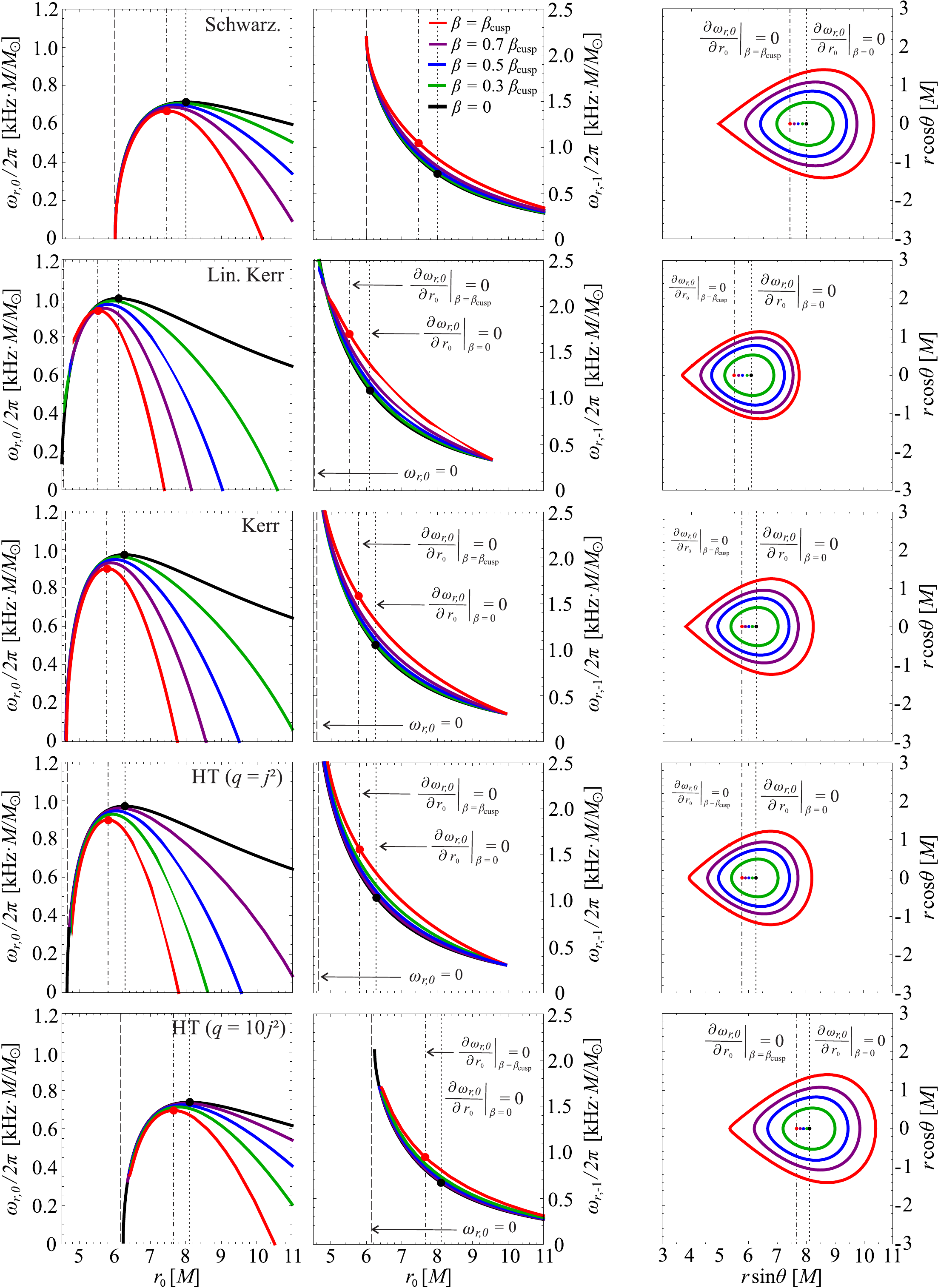}
			\caption{ The same as in Figure \ref{j_0_2} but for $a = 0.4$. \label{j_0_4}}
		\end{center}
	\end{figure}

	We use equation (\ref{radeq}) to derive the radial epicyclic mode frequency as a function of the radius of the torus centre $r_{0}$. In Figure \ref{j_0_2}, we plot the frequencies of both the $m=0$ (left panel) and $m=-1$ (middle panel) radial epicyclic modes. These are compared for the four different geometries assuming $j=0.2$. The right panel of this figure illustrates the behaviour of tori cross-sections corresponding to maxima of the $m=0$ radial epicyclic mode frequency. Figure \ref{j_0_4} then provides the same illustration but for $j=0.4$.
	
	\clearpage
	
	In Table~\ref{tab:geo2}, we provide a quantitative comparison of the maximal frequencies of the $m = 0$ radial epicyclic mode for tori of maximal thicknesses (i.e., the frequencies denoted by the red dots in the left panels of Figures \ref{j_0_2} and \ref{j_0_4}) for the Hartle-Thorne and the other three geometries. In Table~\ref{tab:geo3}, we then present the same but for the $m = -1$ radial epicyclic mode (i.e., the frequencies denoted by the red dots in the middle panels of Figures \ref{j_0_2} and \ref{j_0_4}). The proper radial extension of tori related to Tables~\ref{tab:geo2}~and~\ref{tab:geo3} (i.e., those shown in the right panels of Figures \ref{j_0_2} and \ref{j_0_4}) are compared in Table~\ref{tab:geo4}. 
	
	\begin{table}[ht]
		\caption{The percentual differences in the maximal values of frequencies of the $m = 0$ radial epicyclic mode of the cusp tori in the Hartle-Thorne geometry and in the Schwarzschild, Kerr, and linearized Kerr geometries. The displayed values correspond to the situations illustrated in Figures \ref{j_0_2} and \ref{j_0_4}.}
		\begin{center}
			\renewcommand{\arraystretch}{1.3}
			\begin{tabular}{lcccccc}\hline \hline
				Geometry & \multicolumn{2}{c}{Schwarzschild}   & \multicolumn{2}{c}{Kerr} & \multicolumn{2}{c}{Lin. Kerr}  \\
				\hline
				Spin $j$ & $0.2$ & $0.4$ & $0.2$ & $0.4$ & $0.2$ & $0.4$ \\
				\hline
				HT ($q = j^2$)  & $ +\,15\,\%$ & $ +\,35\,\%$	& $  \,0\,\%$ & $ \,0\,\%$ &  	$- \,1\,\%$ & 	$- \,4\,\%$ \\
				HT ($ q = 10 j^2$) &  $+ \,8\,\%$ & $ +\,4\,\%$& 	 $- \,6\,\%$ & $-\,23\,\%$	& $ -\,7\,\%$ & $- \,26\,\%$  \\
				\hline
				\hline
			\end{tabular}
		\end{center}
		\label{tab:geo2}
	\end{table}
	
	\begin{table}[ht]
		\caption{The percentual differences in the frequency of the $m = -1$ radial epicyclic mode of the cusp tori in the Hartle-Thorne geometry and in the Schwarzschild, Kerr, and linearized Kerr geometries. The frequency is evaluated at the radius at which the $m=0$ radial epicyclic mode frequency has its maximum. The displayed values correspond to the situations illustrated in Figures \ref{j_0_2} and \ref{j_0_4}.}
		
		\begin{center}
			\renewcommand{\arraystretch}{1.3}
			\begin{tabular}{lcccccc}\hline \hline
				Geometry & \multicolumn{2}{c}{Schwarzschild}   & \multicolumn{2}{c}{Kerr} & \multicolumn{2}{c}{Lin. Kerr}  \\
				\hline
				Spin $j$& $0.2$ & $0.4$ & $0.2$ & $0.4$ & $0.2$ & $0.4$ \\
				\hline
				HT ($q = j^2$)  & $ +\,20\,\%$ & $ +\,49\,\%$	& $  \,0\,\%$ & $ -\,1\,\%$ &  	$- \,1\,\%$ & 	$- \,8\,\%$ \\
				HT ($ q = 10 j^2$) &  $+ \,5\,\%$ & $ - \,10\,\%$& 	 $- \,23\,\%$ & $-\,40\,\%$	& $ -\,14\,\%$ & $- \,44\,\%$  \\
				\hline
				\hline
			\end{tabular}
		\end{center}
		\label{tab:geo3}
	\end{table}
	
	\clearpage
	
	\begin{table}[ht]
		\caption{The percentual differences in the proper radial extension $\Delta r_{\mathrm{prop}}$ of the cusp tori relevant to Tables \ref{tab:geo2} and \ref{tab:geo3} and shown in the right panels of Figures \ref{j_0_2} and \ref{j_0_4}.}
		
		\begin{center}
			\renewcommand{\arraystretch}{1.3}
			\begin{tabular}{lcccccc}\hline \hline
				Geometry & \multicolumn{2}{c}{Schwarzschild}   & \multicolumn{2}{c}{Kerr} & \multicolumn{2}{c}{Lin. Kerr}  \\
				\hline
				Spin $j$ & $0.2$ & $0.4$ & $0.2$ & $0.4$ & $0.2$ & $0.4$ \\
				\hline
				HT ($q = j^2$)  & $-\,7\,\%$ & $ -\,17\,\%$	& $  0\,\%$ & $ -\,4\,\%$ &  	$0\,\%$ & 	$+\,1\,\%$ \\
				HT ($ q = 10 j^2$) &  $-\,6\,\%$ & $-\,7\,\%$& 	 $-\,4\,\%$ & $+\,7\,\%$	& $+\,2\,\%$ & $+\,13\,\%$  \\
				\hline
				\hline
			\end{tabular}
		\end{center}
		\label{tab:geo4}
	\end{table}
	
	\begin{figure}[hb!]
		
		\begin{center}
			\includegraphics[width=0.85\linewidth]{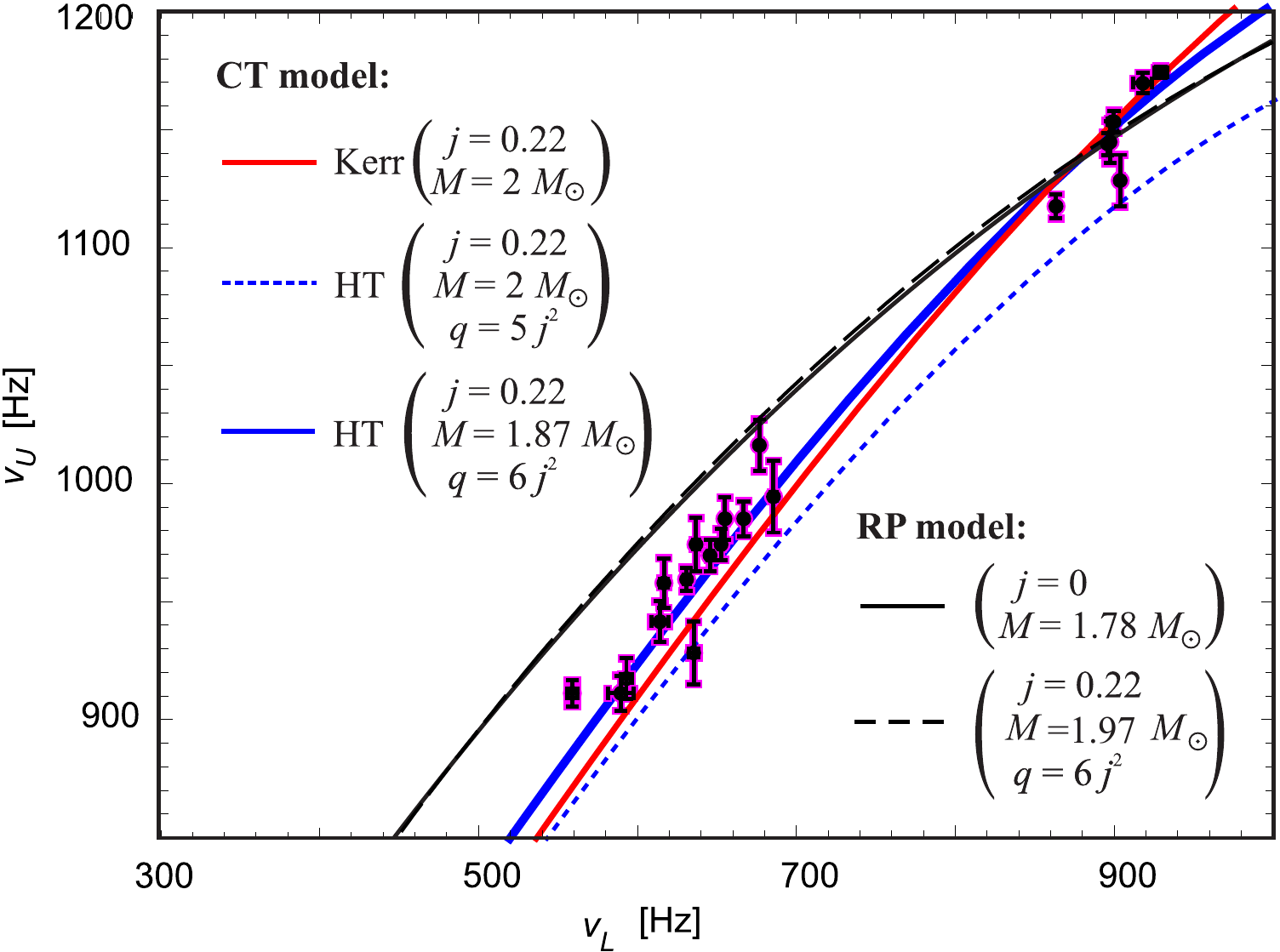}
			\caption{Frequency correlations predicted by the CT model vs. data of the 4U~1636-53 atoll source.  The fit for $j=0.22$ obtained under the consideration of the Kerr geometry (the curve marked as Kerr) is compared here to two examples of predictions obtained under the consideration of the Hartle-Thorne geometry (the curves marked as HT). Examples of the best fits predicted by the relativistic precession model for a given $j$ and $q$ are shown as well (the curves marked as RP model).
				\label{figure:impact}
			}
		\end{center}
	\end{figure}
	
	\section{Discussion and conclusions}
	
	Our results indicate that, while the shape of the non-oscillating tori is not much sensitive to the NS quadrupole moment, the frequencies of the radial epicyclic modes of tori oscillations are affected significantly.
	Clearly, the difference of the frequencies of oscillations of tori around BHs and NSs can reach tens of percents.
	Although a more detailed analysis is certainly needed ( including the completion of the radial epicyclic mode investigation as well as the investigation of the vertical epicyclic mode behaviour), we may already conclude that the consideration of the quadrupole moment induced by the NS rotation likely should have an impact on the modeling of the high-frequency quasi-periodic oscillations.
	
	Our conclusion is demonstrated in Figure \ref{figure:impact}. There we consider a recently proposed QPO model \citep[CT model;][]{tor-etal:2016:MNRAS} and compare the frequencies predicted by the model for several combinations of $M,j,q$ with the frequencies observed in the 4U~1636-53 atoll source  \citep[the data are taken from][]{bar-etal:2006,tor:2009:}. We include in the figure examples of  correlations predicted by the relativistic precession model \citep[][]{ste-etal:1999}. This model provides less promising fits of the data than the CT model while the effects associated to the NS rotation do not imply a significant improvement \citep[see][]{Torok2012,tor-etal:2016:ApJ,tor-etal:2016:MNRAS}. It is clear from the figure that even when we restrict ourselves to values of the Hartle-Thorne spacetime parameters that are consistent with up-to-date models of neutron stars, no conceivable smooth curve can reproduce the data in a significantly better way compared to the CT model.

	
	
	\ack
	We acknowledge two internal grants of the Silesian University, $\mathrm{SGS/12,13/2019}$. We wish to thank the INTER-EXCELLENCE project No. $\mathrm{LTI17018}$.  KK thanks to the INTER-EXCELLENCE project No. LTT17003.  DL thanks the Student Grant Foundation of the Silesian University in Opava, Grant No. $\mathrm{SGF/1/2020}$, which has been carried out within the EU OPSRE project entitled ``Improving the quality of the internal grant scheme of the Silesian University in Opava'', reg. number: $\mathrm{CZ.02.2.69/0.0/0.0/19\_073/0016951}$. The autors were also supported by the ESF projects No. $\mathrm{CZ.02.2.69/0.0/0.0/18\_054/0014696}$.
	

\def\prc{Phys. Rev. C}
\def\pre{Phys. Rev. E}
\def\prd{Phys. Rev. D}
\def\jcap{Journal of Cosmology and Astroparticle Physics}
\def\apss{Astrophysics and Space Science}
\def\mnras{Monthly Notices of the Royal Astronomical Society}
\def\apj{The Astrophysical Journal}
\def\aap{Astronomy and Astrophysics}
\def\actaa{Acta Astronomica}
\def\pasj{Publications of the Astronomical Society of Japan}
\def\apjl{Astrophysical Journal Letters}
\def\pasa{Publications Astronomical Society of Australia}
\def\nat{Nature}
\def\physrep{Physics Reports}
\def\araa{Annual Review of Astronomy and Astrophysics}
\def\apjs{The Astrophysical Journal Supplement}
\def\aapr{The Astronomy and Astrophysics Review}
\def\procspie{Proceedings of the SPIE}

\def\pasp{PASP}
\def\aj{AJ}
\def\nat{Nature}
\def\nar{NewAR}
\def\na{NewA}
\def\icarus{Icar}
\def\araa{ARA\&A}
\def\aplett{Astrophysical Letters}
\def\prl{Physical Review Letters}

\def\prc{Phys. Rev. C}
\def\pre{Phys. Rev. E}
\def\prd{Phys. Rev. D}
\def\jcap{Journal of Cosmology and Astroparticle Physics}
\def\apss{Astrophysics and Space Science}
\def\mnras{Mon. Not. R. Astron Soc.}
\def\apj{The Astrophysical Journal}
\def\aap{Astron. Astrophys.}
\def\actaa{Acta Astronomica}
\def\pasj{Publications of the Astronomical Society of Japan}
\def\apjl{Astrophysical Journal Letters}
\def\pasa{Publications Astronomical Society of Australia}
\def\nat{Nature}
\def\physrep{Phys. Rep.}
\def\araa{Annu. Rev. Astron. Astrophys.}
\def\apjs{The Astrophysical Journal Supplement}
\def\aapr{The Astronomy and Astrophysics Review}

\def\mdash{---}
	\bibliographystyle{ragtime}
	\bibliography{mat}
	
\end{document}